\documentclass[runningheads]{llncs}
\usepackage{graphicx}

\usepackage{multirow}
\begin{document}
\title{Cascaded Convolutional Neural Network for Automatic Myocardial Infarction Segmentation from Delayed-Enhancement Cardiac MRI}
\titlerunning{Cascaded CNN for Myocardial Infarction Segmentation}
%
\author{Yichi Zhang }
\authorrunning{Y. Zhang}
%
\institute{School of Biological Science and Medical Engineering, Beihang University}

\maketitle              

\begin{abstract}
Automatic segmentation of myocardial contours and relevant areas like infraction and no-reflow is an important step for the quantitative evaluation of myocardial infarction. In this work, we propose a cascaded convolutional neural network for automatic myocardial infarction segmentation from delayed-enhancement cardiac MRI. We first use a 2D U-Net to focus on the intra-slice information to perform a preliminary segmentation. After that, we use a 3D U-Net to utilize the volumetric spatial information for a subtle segmentation. Our method is evaluated on the MICCAI 2020 EMIDEC challenge dataset and achieves average Dice score of 0.8786, 0.7124 and 0.7851 for myocardium, infarction and no-reflow respectively, outperforms all the other teams of the segmentation contest.

\end{abstract}
\section{Introduction}
Myocardial infarction (MI) is a myocardial ischemic necrosis caused by coronary artery complications that cannot provide enough blood and has become one of the leading causes of death and disability worldwide \cite{thygesen2007universal}. The viability of the cardiac segment is an important parameter to assess the cardiac status after MI, such as whether the segment is functional after the revascularization. Delayed-enhancement MRI (DE-MRI) performed several minutes after the injection is a method to evaluate the extent of MI and assess viable tissues after the injury. According to the World Health Organization (WHO), cardiovascular diseases are the first one cause of death worldwide and 85\% of the deaths are due to heart attacks and strokes. 

Automatic segmentation of the different relevant areas from DE-MRI, such as myocardial contours, the infarcted area and the permanent microvascular obstruction area (no-reflow area) could provide useful information like the absolute value (mm$^3$) or percentage of the myocardium, which can provide useful information for quantitative evaluation of MI. However, myocardial infarction segmentation is still a challenging task due to the morphological similarity. Recently, deep learning-based methods have achieved state-of-the-art results for various image segmentation tasks and shown great potential in medical image analysis and clinical applications. 

In this paper, we propose a cascaded convolutional neural network for automatic myocardial infarction segmentation from delayed-enhancement cardiac MRI. We first use a 2D U-Net to focus on the intra-slice information and perform a preliminary segmentation. After that, a 3D U-Net is applied to utilize the volumetric spatial information for a subtle segmentation. All the training procedure of our network are based on MICCAI 2020 EMIDEC Challenge dataset\footnote{http://emidec.com} \cite{lalande2020emidec} .

\section{Method}

\subsubsection{Network Architecture} 

As the most well-known network structure for medical image segmentation, U-Net \cite{ronneberger2015u} is a classical encoder-decoder segmentation network and achieves state-of-the-art results on many segmentation challenges \cite{heller2020state,isensee2019nnu}. The encoder is similar with the typical classification network and uses convolution-pooling module to extract more high-level semantic features layer by layer. Then the decoder recovers the localization for every voxel and utilizes the extracted feature information for the classification of each pixel. To incorporate multi-scale features and employ the position information, skip connections are constructed between the encoder and decoder in the same stage.

For the segmentation of 3D biomedical images, many 3D segmentation networks like \cite{cciccek20163d,milletari2016v} are proposed to extract volumetric spatial information using 3D convolutions instead of just focusing on intra-slice information. However, for some volumes with highly anisotropic voxel spacings, 3D networks may not always outperform 2D networks when the inter-slice correlation information is not rich \cite{abulnaga2018ischemic}. For example, Case N042 is a 3D MRI volume with an image shape of 166*270*7 and voxel spacing of 1.667mm*1.667mm*10mm on x, y, and z-axis, respectively. This means x and y-axis preserve much higher resolution and richer information than the z-axis. Under this circumstance, using pure 3D network that treats the three axes equally may not be the best choice.

\begin{figure}[!htb]
	\includegraphics[width=12cm]{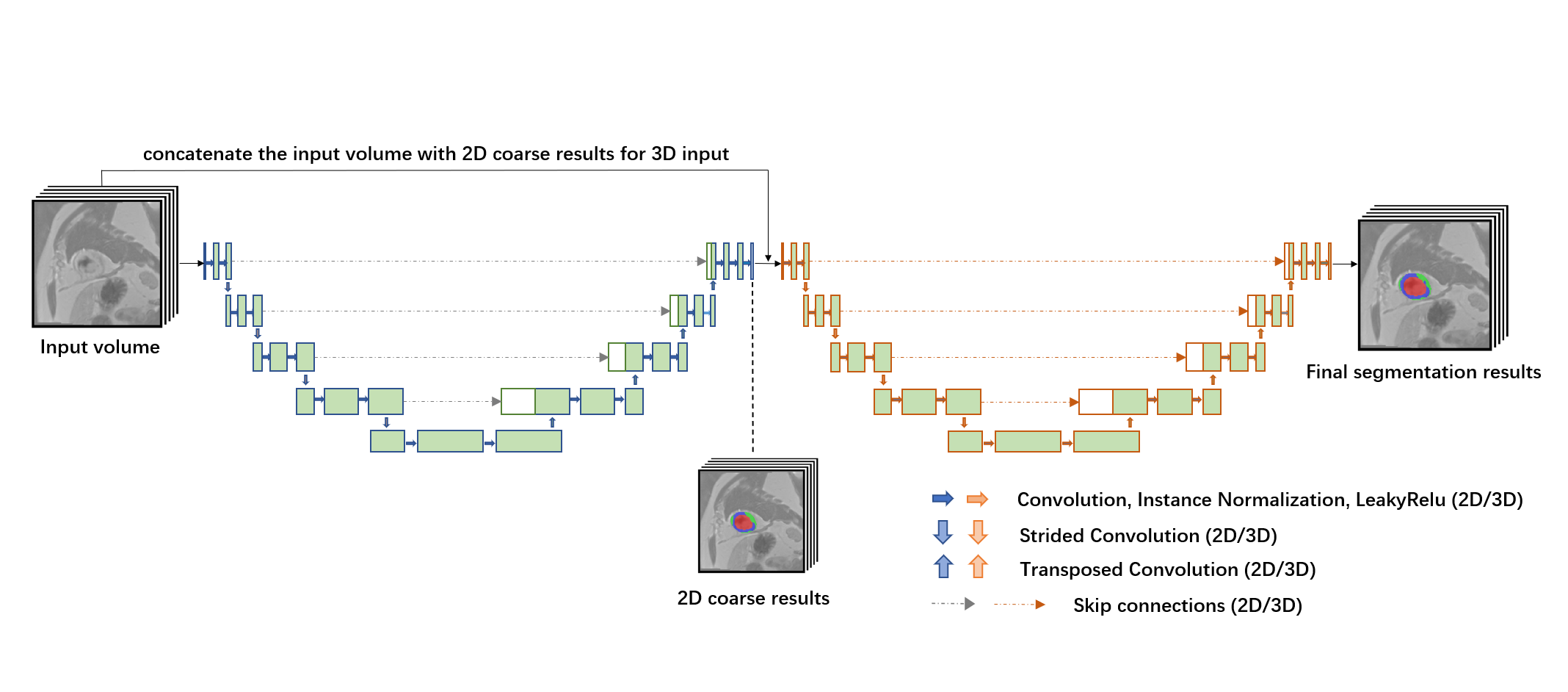}
	\centering
	\caption{The overall architecture of our cascaded convolutional neural network. The blocks and arrows viewed in blue and red denote the corresponding structure for 2D and 3D networks. }
	\label{Fig1}
\end{figure}

To issue this problem, we propose a cascaded convolutional neural network for automatic myocardial infarction segmentation from delayed-enhancement cardiac MRI. As illustrated in Fig.\ref{Fig1}, our network can be mainly divided into two stages. Firstly, after the preprocessing of input data, the MRI volume is divided into a sequence of slices for input of 2D U-Net to obtain a preliminary segmentation based on the intra-slice information. However, the results of 2D network ignore the correlation between slices, which would lead to limited segmentation accuracy, especially for challenging pathological areas that are difficult to distinguish only based on intra-slice information. Therefore, in the second stage, we use a 3D U-Net to utilize the volumetric spatial information and make a subtle segmentation. Specifically, we concatenate the 2D coarse results in the first stage with the input volume for the input of 3D U-Net in the second stage as a spatial prior. In the end, after the postprocessing like removing the scattered voxels, we get the final segmentation results.

\subsubsection{Implementation Details} 

All the training procedure of our network is performed on NVIDIA Tesla V100 GPUs using the Pytorch framework based on the nnU-Net implementation \cite{isensee2019nnu}. During training, we use Adam optimizer with an initial learning rate of 0.01. Instead of patch-based methods, we use the whole short-axis slice and whole volume for the input of the 2D and 3D networks. To enhance the attention of foreground voxels, we use the summation of cross-entropy (CE) loss and dice loss \cite{milletari2016v} as the loss function for the training of our network.

\subsubsection{Dataset and Evaluation Metrics}

The EMIDEC Challenge dataset consists of delayed-enhancement cardiac MRI with a training set of 100 patients including 67 pathological cases and 33 normal cases, and a testing set of another 50 patients including 33 pathological cases and 17 normal cases. For training cases, manual annotations are provided with 0 for background, 1 for cavity, 2 for normal myocardium, 3 for myocardial infarction, and 4 for no-reflow. 

For evaluation of segmentation results, clinical metrics include the average errors for the volume of the myocardium of the left ventricle, the volume and the percentage of MI and no-reflow and geometrical metrics include the average Dice coefficient for the different areas and Hausdorff distance for the myocardium.

\section{Experiments and Results}

There are totally 100 scans with published labels to train our network, while the other 50 scans remained for final evaluation. We make random 5-fold cross validation by randomly shuffling the sequence of cases and splitting the training dataset into 5 fixed folds with 20 MR scans in each fold, using 4 folds for training and the other one for testing. In this way, we can make a more comprehensive evaluation of our method.

Table \ref{Table1} and Table \ref{Table2} respectively represent the cross-validation results of our 2D coarse segmentation output and final segmentation output. The evaluation of clinical and geometrical metrics is based on the official code\footnote{https://github.com/EMIDEC-Challenge/Evaluation-metrics}. From the result, we can see that the application of 3D U-Net can make use of the volumetric spatial information and improve the segmentation result.

\begin{table}[]
	\caption{Quantitative 5-fold cross-validation results of 2D coarse segmentation output.} \label{Table1}
	\centering
	\setlength{\tabcolsep}{2.5mm}
	\renewcommand\arraystretch{1.1}
	\begin{tabular}{ccccccc}
		\hline
		Targets                     & Metrics     & fold   0 & fold   1 & fold   2 & fold   3 & fold   4 \\ \hline
		\multirow{3}{*}{Myocardium} & Dice(\%)    & 83.98    & 85.29    & 85.94    & 85.83    & 85.59    \\
		& VolDif(mm$^3$) & 10906.43 & 6384.88  & 6012.93  & 5423.57  & 11629.66 \\
		& HSD(mm)     & 17.01    & 13.77    & 13.68    & 12.60    & 12.65    \\ \hline
		\multirow{3}{*}{Infarction} & Dice(\%)    & 44.39    & 53.12    & 48.10    & 50.55    & 66.34    \\
		& VolDif(mm$^3$) & 9883.17  & 4821.30  & 3449.09  & 4986.66  & 6621.52  \\
		& Ratio(\%)   & 7.10     & 4.39     & 2.96     & 4.68     & 5.00     \\ \hline
		\multirow{3}{*}{NoReflow}   & Dice(\%)    & 65.26    & 63.84    & 70.61    & 60.24    & 66.67    \\
		& VolDif(mm$^3$) & 2703.34  & 775.07   & 480.32   & 703.7    & 443.86   \\
		& Ratio(\%)   & 1.68     & 0.68     & 0.37     & 0.67     & 0.32    \\ \hline
	\end{tabular}
\end{table}

\begin{table}[]
	\caption{Quantitative 5-fold cross-validation results of our final segmentation output.} \label{Table2}
	\centering
	\setlength{\tabcolsep}{2.5mm}
	\renewcommand\arraystretch{1.1}
	\begin{tabular}{ccccccc}
		\hline
		Targets                     & Metrics     & fold   0 & fold   1 & fold   2 & fold   3 & fold   4 \\ \hline
		\multirow{3}{*}{Myocardium} & Dice(\%)    & 86.66    & 86.46    & 87.87    & 87.61    & 87.13    \\
		& VolDif(mm$^3$) & 8680.23  & 5405.52  & 6087.88  & 4880.5   & 7317.76  \\
		& HSD(mm)     & 15.88    & 14.12    & 12.96    & 13.43    & 13.79    \\ \hline
		\multirow{3}{*}{Infarction} & Dice(\%)    & 61.44    & 72.08    & 81.51    & 68.48    & 76.87    \\
		& VolDif(mm$^3$) & 6536.55  & 3233.94  & 3514.97  & 4091.74  & 3520.3   \\
		& Ratio(\%)   & 4.67     & 2.91     & 2.85     & 3.96     & 2.64     \\ \hline
		\multirow{3}{*}{NoReflow}   & Dice(\%)    & 68.47    & 68.33    & 79.67    & 65.12    & 73.48    \\
		& VolDif(mm$^3$) & 2158.34  & 712.36   & 451.84   & 620.93   & 649.98   \\
		& Ratio(\%)   & 1.37     & 0.65     & 0.35     & 0.61     & 0.46    \\ \hline
	\end{tabular}
\end{table}

\begin{figure}[!htb]
	\includegraphics[width=12cm]{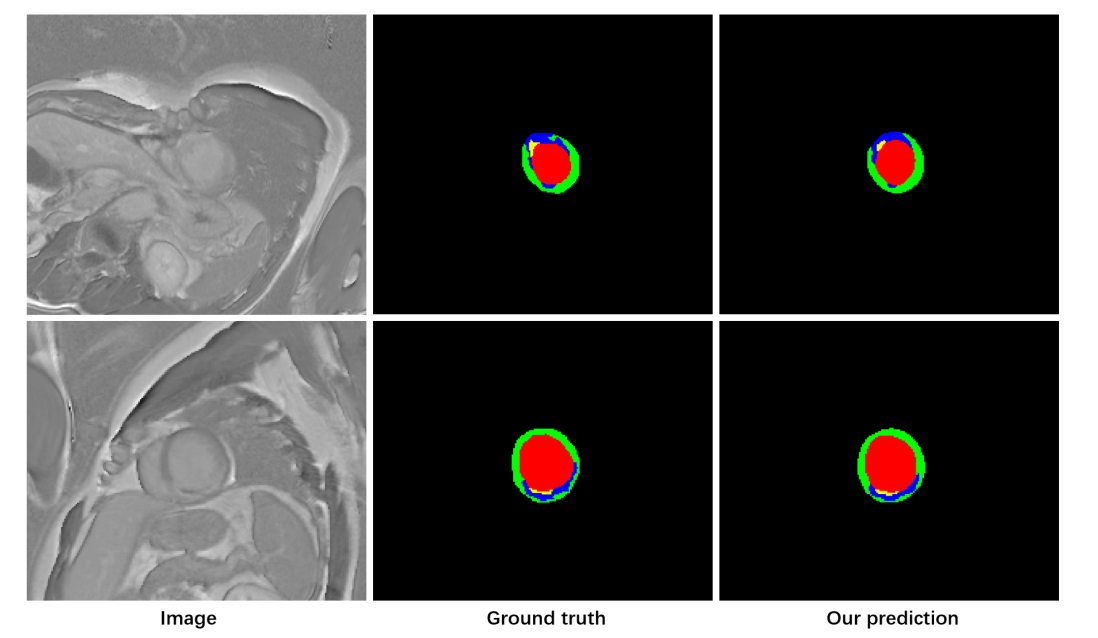}
	\centering
	\caption{Two samples of our segmentation results. The columns from left to right are the image, ground truth, and prediction. (cavity in red, myocardium in green, infarction in blue, no-reflow in yellow)}
	\label{Fig2}
\end{figure}

For our final segmentation results, the network performs well on myocardium segmentation, with an average dice score of 0.8715. However, for more challenging segmentation of pathological areas, the average dice score is only 0.7208 and 0.7101 for infarction and no-reflow. Also, the performance variance is very small, which indicates the robustness of our method. Fig.\ref{Fig2} illustrates two samples of our segmentation results and corresponding ground truth in the validation set of our own split. We can see that the segmentation results closely approximate the ground truth.

In the inference stage, we obtain the final prediction of the testing set by ensembling the segmentation results of each fold using majority voting. The evaluation results of our method on the testing set of EMIDEC dataset is presented in Table \ref{Table3}. The average Dice score is very similar to our cross-validation results (even higher on some metrics), which indicates that our method is stable for the myocardial infarction segmentation task.

\begin{table}[]
	\caption{The Evaluation results of our method on the EMIDEC test set.} \label{Table3}
	\setlength{\tabcolsep}{4mm}
	\renewcommand\arraystretch{1.1}
	\begin{tabular}{ccccc}
		\hline
		Targets	& Dice(\%) & VolDif(mm$^3$) & HSD(mm) & ratio(\%) \\ \hline
		Myocardium & 87.86    & 9258.24     & 13.01   & -         \\
		Infarction & 71.24    & 3117.88     & -       & 2.38      \\
		NoReflow   & 78.51    & 634.69      & -       & 0.38     \\ \hline
	\end{tabular}
\end{table}

\section{Conclusion}

In this paper, we propose a cascaded convolutional neural network for automatic myocardial infarction segmentation from delayed-enhancement cardiac MRI. The network consists of a 2D U-Net to focus on the intra-slice information to perform a preliminary segmentation and a 3D U-Net to utilize the volumetric spatial information to make a subtle segmentation. Our method is trained and validated on MICCAI 2020 EMIDEC challenge dataset. For the testing stage, our ensembled model has achieved an average Dice score of 0.8786, 0.7124 and 0.7851 for myocardium, ranking first of the segmentation challenge.

%
%
%
\bibliographystyle{splncs04}
\bibliography{bibliography}

\end{document}